# Heterogeneous V2V Communications in Multi-Link and Multi-RAT Vehicular Networks

Miguel Sepulcre and Javier Gozalvez

**Abstract**—Connected and automated vehicles will enable advanced traffic safety and efficiency applications thanks to the dynamic exchange of information between vehicles, and between vehicles and infrastructure nodes. Connected vehicles can utilize IEEE 802.11p for vehicle-to-vehicle (V2V) and vehicle-to-infrastructure (V2I) communications. However, a widespread deployment of connected vehicles and the introduction of connected automated driving applications will notably increase the bandwidth and scalability requirements of vehicular networks. This paper proposes to address these challenges through the adoption of heterogeneous V2V communications in multi-link and multi-RAT vehicular networks. In particular, the paper proposes the first distributed (and decentralized) context-aware heterogeneous V2V communications algorithm that is technology and application agnostic, and that allows each vehicle to autonomously and dynamically select its communications technology taking into account its application requirements and the communication context conditions. This study demonstrates the potential of heterogeneous V2V communications, and the capability of the proposed algorithm to satisfy the vehicles' application requirements while approaching the estimated upper bound network capacity.

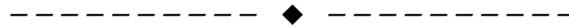

## 1 INTRODUCTION

Connected vehicles will improve traffic safety and efficiency thanks to the wireless exchange of information between vehicles (Vehicle-to-Vehicle or V2V communications), and between vehicles and infrastructure nodes (Vehicle-to-Infrastructure or V2I communications). Cooperative active safety applications (e.g. emergency electronic brake lights, intersection collision avoidance or lane change warning) generally require the periodic transmission and reception of broadcast messages that include basic positioning and status information; these messages are known as CAMs (Cooperative Awareness Messages) in Europe and BSMs (Basic Safety Messages) in the US. They can be transmitted using IEEE 802.11p, also known as ITS-G5 (Intelligent Transportation Systems-G5) in Europe and DSRC (Dedicated Short-Range Communications) in the US [1].

The introduction of connected automated vehicles will increase the reliability, latency and bandwidth requirements of vehicular communications [2]. Connected automated vehicles will benefit from the implementation of cooperative driving maneuvers where nearby vehicles exchange information to safely coordinate driving maneuvers such as entering a roundabout/highway or changing lanes. This exchange requires very reliable and low latency V2V communications. Also, the exchange of rich sensor data between vehicles can improve their capacity to collaboratively detect, estimate and characterize the local environment (referred to as collective perception or cooperative sensing). Exchanging this information can require large communication bandwidths. A connected vehicle transmitting CAMs/BSMs (~200Bytes) at 10Hz requires a communications link of ~16Kbps. However, the throughput required by connected automated vehicles can be in the order of Mbps [3][4], which results in more stringent requirements in terms of bandwidth.

An approach to support connected automated vehicles and its higher communication requirements is the development of heterogeneous V2X communications and networks. Heterogeneous wireless networking has been utilized in cellular networks to increase the communication bandwidth and improve the networks' scalability [5]. Cooperative ITS standards for V2X (Vehicle-to-Everything) communications allow for the implementation of heterogeneous vehicular communications. For example, the ITS station reference architecture standardized by ISO (International Organization for Standardization) [6] considers the possibility to use different Radio Access Technologies (RATs) at the physical and MAC (Medium Access Control) layers. This architecture has been adapted to the European context by ETSI (European Telecommunications Standards Institute) [7]. ETSI currently runs two study items to investigate further enhancements to this architecture in order to support communications between vehicles with multiple RATs [8][9]. The active study items are currently analyzing different implementation and deployment options including a multi-link and multi-RAT scenario where all vehicles can simultaneously receive messages using different RATs. Multi-link is the capability of a device to communicate using multiple wireless links simultaneously. Multi-RAT is the capability of a device to communicate using different RATs. However, a device cannot transmit using multiple RATs simultaneously unless it also implements multi-link capabilities. In a multi-RAT scenario, devices im-

_________________
- *Miguel Sepulcre and Javier Gozalvez are with Universidad Miguel Hernandez de Elche (UMH), Avda. Universidad s/n, 03202, Elche (Alicante), Spain. E-mail: msepulcre@umh.es, j.gozalvez@umh.es.*

plement multiple RATs, but they cannot use them at the same time. Devices must select at each point in time the RAT they would like to utilize to transmit data and receive data from other vehicles. In a multi-link and multi-RAT scenario, vehicles can dynamically select the RAT they use to transmit while using simultaneously the other RATs to receive information from other vehicles. 3GPP (3rd Generation Partnership Project) also considers the use of multiple RATs to support the 5G eV2X applications (including autonomous driving) [4]. 5G-PPP (5G Infrastructure Public Private Partnership) also highlights multi-link and multi-RAT connectivity as a promising approach to support future automotive use cases [2].

Standards have defined the main components needed for the implementation of heterogeneous V2X communications, but do not define specific heterogeneous V2X algorithms. To date, heterogeneous vehicular networking has been mainly applied to V2I communications (e.g. [10], [11]) since most current bandwidth-demanding applications are Internet-based and require the connection to the infrastructure. However, V2V communications will also be challenged (both in terms of reliability and bandwidth) under dense deployment scenarios, and with the introduction of connected automated vehicles that will have higher bandwidth demands. In this context, this paper proposes to exploit heterogeneous V2V communications to support connected and automated vehicles. To this aim, the paper presents CARHet (Context-AwaRe Heterogeneous V2V communications), the first decentralized heterogeneous V2V communications algorithm for multi-link and multi-RAT vehicular networks that is technology and application agnostic. CARHet allows each vehicle to dynamically select its radio access technology taking into account its application requirements and the communication context conditions observed by other vehicles (e.g. the channel load level they have measured for each RAT). The conducted evaluation demonstrates the potential of heterogeneous V2V communications, and the capacity of the proposed algorithm to satisfy the application requirements while approaching the estimated upper bound of the vehicular network capacity with a low computational cost and communications overhead.

## 2 STATE OF THE ART

Heterogeneous networking has been largely investigated in the context of cellular systems. In cellular systems, the core network selects the most suitable RAT for each device. The selection usually takes into account context information available at the core network and obtained from the devices. Several studies have demonstrated the significant gains that heterogeneous networking can provide, e.g. higher bit rates or network capacity [12].

Several studies have highlighted the benefits of applying heterogeneous networking to V2I communications [13], and first algorithms to select the most adequate V2I communications technology at each point in time have been proposed in the literature. For example, [10] proposes a method to select the communications technology (WiFi or LTE in their study) that maximizes the QoE (Quality of Experience) during a vehicle's route. The method takes into account the service type, the vehicle's route, and the traffic dynamics over the backhaul links of each technology. The selection algorithm proposed in [11] takes into account user preferences, and selects the technology that better fulfils the application requirements. The algorithm exploits location and navigation information to minimize the number of handovers between technologies during the vehicle's route. The algorithm presented in [14] focuses on the interworking of cellular and WiFi networks. The study concludes that it is possible to minimize the transmission time if vehicles switch from cellular to WiFi when approaching WiFi access points, but only when vehicles move at low speeds. The authors presented in [15] a network-assisted heterogeneous V2I algorithm designed to improve both the individual and system performance. The selection process takes into account context information such as the position of base stations, the vehicle's route and the travel time.

Some similarities exist between heterogeneous networking and Dynamic Spectrum Access (DSA), although they differ on the problem addressed and their objectives. DSA considers that primary and secondary users can share a given spectrum band [16]. In particular, secondary users are allowed to use the band if primary users are not using it. To this aim, secondary users utilize cognitive radios and must sense the spectrum band in order to detect the potential presence of primary users. However, the radios implement a single RAT and the objective is to efficiently and reliably find transmission opportunities in a spectrum band that is primarily assigned to other users. DSA has been applied to vehicular networks e.g. using the TV white space band [17] or exploiting historical spectrum sensing data [18]. In heterogeneous networking, all users are considered primary users and they are all equipped with multiple RATs that are assigned specific channels and spectrum bands. In this case, the objective is for each user to select the most adequate RAT based on its application requirements and context conditions.

Limited work has been done to date to apply heterogeneous networking to V2V communications. This is partly due to the fact that IEEE 802.11p has generally been considered as the de-facto technology for V2V communications. However, the limitations of IEEE 802.11p and the emergence of other device-to-device technologies (e.g. LTE-V [19], WiFi-Direct [20] or even Visible Light Communications [21]) paves the way for applying heterogeneous networking to V2V communications in order to improve the reliability, bandwidth and scalability of vehicular networks. It is important noting that the algorithms and conclusions derived from heterogeneous V2I studies cannot be directly applied to heterogeneous V2V communications. This is the case because the communication requirements are different, and also because many of the assumptions made for V2I scenarios are not valid, e.g. the static position of the target communicating nodes.

First studies considering the use of different RATs for V2V communications have proposed the use of cellular technologies as a backup when IEEE 802.11p-based V2V multi-hop connections cannot be established [22]. For example, [23] suggests using cellular D2D (Device-to-Device) communications as a failover recovery solution in multi-hop V2V connections. Conventional infrastructure-based cellular communications have also been proposed to improve the V2V connectivity in the case of low IEEE 802.11p penetration rates. For example, [24] proposes an application layer handoff that simultaneously transmits event-driven messages through IEEE 802.11p (V2V) and LTE (V2I-I2V) in order to disseminate safety-critical collision warning messages to nearby vehicles. Similarly, [25] proposes a hybrid architecture for safety message dissemination that organizes the IEEE 802.11p network in clusters using V2V communications. Cluster heads operate using dual radio interfaces in order to connect the IEEE 802.11p sub-networks to the LTE network. [26] is one of the first studies that has proposed using different V2V communication technologies for connected automated vehicular applications, in particular to manage platoons. The study proposes that only platoon leaders should use IEEE 802.11p while the following vehicles in a platoon should communicate using Visible Light Communications. The objective is to improve the reliability and scalability of the network by reducing the use of IEEE 802.11p.

Existing studies have provided first insights into the potential of applying heterogeneous networking concepts to V2V communications. These studies have proposed policies to decide when each communication technology should be utilized. This paper complements the existing state of the art by presenting what is, to the authors' knowledge, the first heterogeneous V2V communications algorithm for multi-link and multi-RAT vehicular networks that is technology and application agnostic. The proposed algorithm allows each vehicle to autonomously and dynamically select its V2V communications technology (with a low computational cost and overhead) based on its application requirements and the communication context conditions observed by its 1-hop neighboring vehicles. The use of multiple RATs in multi-link scenarios allows exploiting the different characteristics of each RAT and their complementarities (e.g. the use of different bandwidth and spectrum bands, medium access control and physical layer schemes, etc.). This is not possible in multi-link scenarios where vehicles have multiple wireless links, but only implement a single RAT.

## 3 HETEROGENEOUS V2V COMMUNICATIONS: FRAMEWORK AND MOTIVATION

This study proposes the use of heterogeneous V2V communications in order to help address the bandwidth demands of future connected automated vehicles, and support the implementation of cooperative perception and driving applications. To this aim, we propose a distributed heterogeneous V2V communications algorithm that allows each vehicle to dynamically select the RAT that is more suitable at each point in time. This study considers a multi-link and multi-RAT vehicular scenario where all vehicles are equipped with different RATs operating in different bands[1]. In line with 5G-PPP [2], 3GPP [4] and ETSI [9], we consider that vehicles are able to simultaneously use different RATs for data transmission and/or reception. A vehicle can then transmit data using one RAT and simultaneously receive information through all available RATs as shown in Fig. 1.

This section illustrates the potential of heterogeneous V2V communications to increase the network capacity. To this aim, the section assumes that all vehicles have the same bandwidth demand, and compares the upper-bound of the traffic density that could be supported when using a single RAT per vehicle, and when implementing heterogeneous V2V communications at each vehicle. The maximum traffic density can be estimated as a function of the channel load. The channel load is typically measured using the CBR (Channel Busy Ratio) metric, which represents the percentage of time that the channel is sensed as busy. The CBR experienced at a given position $x$ when using a radio access technology $r$ can be estimated as the summation of the load contribution of all the vehicles in the scenario that also transmit using technology $r$:

$$CBR_r(x) = \sum_{i=1}^{N_r} n_i \cdot t_i \cdot PSR_r(|x_i - x|) \tag{1}$$

where $x_i$ represents the position of vehicle $i$, $n_i$ the number of packets vehicle $i$ transmits per second, $t_i$ the time duration of each transmitted packet and $N_r$ is the number of vehicles transmitting using technology $r$. PSR (Packet Sensing Ratio) is a distance-dependent function that represents the probability that a packet is sensed at a given distance to the transmitter. The PSR function depends on different factors such as the transmission power, the radio propagation conditions and the carrier sense threshold. Without loss of generality, if we consider that all vehicles are uniformly distributed (with an inter-vehicle distance of $d$), and they all transmit the same number of packets per second ($n_{pkt} = n_i$) with the same duration ($t_{pkt} = t_i$), equation (1) can be transformed for $x=0$ into:

$$CBR_r = n_{pkt} \cdot t_{pkt} \cdot \beta \cdot \sum_{i=1}^{N_r} PSR_r(|i|) \tag{2}$$

where $\beta=1/d$ represents the vehicle density and is expressed in vehicles per meter if the distance between vehicles ($d$) is expressed in meters. The same results would be obtained for other values of $x$. If we consider that the maximum CBR that can be experienced with a given radio access technology $r$ is $CBR_r^{max}$, then the maximum traffic density that could be supported by this technology is:

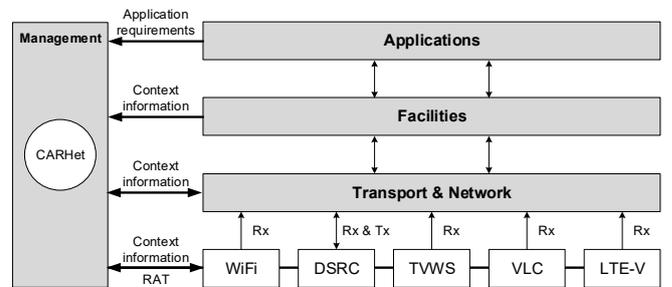

Fig. 1. Architecture and concept illustration of heterogeneous V2V communications. DSRC is used for data transmission and reception, while all other RATs are used for receiving data only.

[1] Each RAT utilizes a single and pre-defined channel.

$$\beta_r^{\max} = \frac{CBR_r^{\max}}{n_{pkt} \cdot t_{pkt} \cdot \sum_{i=1}^{N_r} PSR_r(|i|)} \quad (3)$$

The maximum traffic density that can be supported increases with the number of RATs available at each vehicle to transmit data. Let's consider that each vehicle has $N_{RAT}$ radio access technologies. If data transmissions are adequately distributed over the different RATs, the maximum traffic density that can be supported when implementing heterogeneous V2V communications can be approximated by:

$$\beta^{\max} = \sum_{r=1}^{N_{RAT}} \beta_r^{\max} \quad (4)$$

Without loss of generality, this work considers that each vehicle is equipped with 5 RATs: DSRC (IEEE 802.11p) operating at 5.9GHz, DSRC (IEEE 802.11p) operating at 700MHz, WiFi operating at 5.6GHz, WiFi operating at 2.4GHz, and an OFDM-like technology operating in the TVWS (TV White Space) band at 460MHz. Table I reports the main communication parameters for each RAT. These parameters are fixed in this study since our objective is not to optimize the operation of each RAT, but instead illustrate the potential of heterogeneous V2V communications. The minimum signal level needed to correctly receive a packet (i.e. the reception threshold) has been set 3dB higher than the noise power for all RATs. The transmission power levels have been configured to the maximum values for each RAT. The propagation conditions are modeled using the Winner+ B1 propagation model recommended by the EU project METIS (Mobile and wireless communications Enablers for the Twenty-twenty Information Society) for D2D/V2V [27]. This model is valid for the frequency range 0.45-6GHz. Winner+ B1 includes a log-distance pathloss model for the average propagation loss as a function of the distance between transmitter and receiver. A log-normal random variable is used to model the shadowing effect caused by surrounding obstacles. The model differentiates between LOS (Line-of-Sight) and Non-LOS conditions. Using the Winner+ B1 model and the parameters in Table I, we have derived the PSR curves for each RAT that are needed to estimate the maximum traffic density supported by heterogeneous V2V communications. The PSR curves are shown in Fig. 2.

Fig. 3 compares the maximum traffic density ($\beta^{max}$) that could be supported with heterogeneous V2V communications and with each one of the 5 available RATs when utilized individually. Fig. 3 has been obtained using eq. (4) and Fig. 2, and setting the maximum CBR for all RATs to 0.6 [28]. Fig. 3a depicts the results considering the same MCS (Modulation and Coding Scheme) - QPSK ½ - for all RATs. This MCS corresponds to a data rate of 6Mbps for IEEE 802.11p at 5.9GHz, which is the default MCS proposed by the ETSI standards. Fig. 3b plots the results using the highest MCS for each RAT (i.e. the data rates shown in Table I); increasing the data rate augments the maximum traffic density. Independently of the MCS utilized, Fig. 3 clearly illustrates the capacity gains that can be obtained with heterogeneous V2V communications. In all the scenarios considered, using heterogeneous V2V communications could increase the capacity by approximately 8x compared to when using only IEEE 802.11p at 5.9GHz. For example, in a scenario with all vehicles transmitting 0.5Mbps and using the highest MCS, IEEE 802.11p at 5.9GHz (DSRC59) could support only 35 vehicles/km. This number can increase to up to 280 vehicles/km when using heterogeneous V2V communications. It is also interesting to note that the same gain is achieved when considering connected vehicles transmitting CAMs/BSMs (i.e. around 200Bytes at 10Hz or 16Kbps). In this case, the estimated maximum traffic density supported by DSRC59 would be 265 vehicles/km (with QPSK ½). This value could increase to more than 2200 vehicles/km with heterogeneous V2V communications. These results illustrate the capacity gains that can be achieved with heterogeneous V2V communications. Achieving such gains requires the design of an heterogeneous V2V communications algorithm that distributes the transmissions over the different RATs. This is the objective of CARHet that is presented in the next section.

TABLE I. COMMUNICATION PARAMETERS

| Parameter | DSRC 0.7GHz | DSRC 5.9GHz | WiFi 2.4GHz | WiFi 5.6GHz | TVWS |
|---|---|---|---|---|---|
| Carrier freq. [GHz] | 0.7 | 5.9 | 2.4 | 5.6 | 0.46 |
| Bandwidth [MHz] | 10 | 10 | 20 | 20 | 6 |
| Tx. power [dBm] | 10 | 23 | 20 | 17 | 20 |
| Noise [dBm] | -97 | -97 | -94 | -94 | -99 |
| Rx. Threshold [dBm] | -94 | -94 | -91 | -91 | -96 |
| Data rate [Mbps] | 18 | 27 | 54 | 54 | 7.2 |

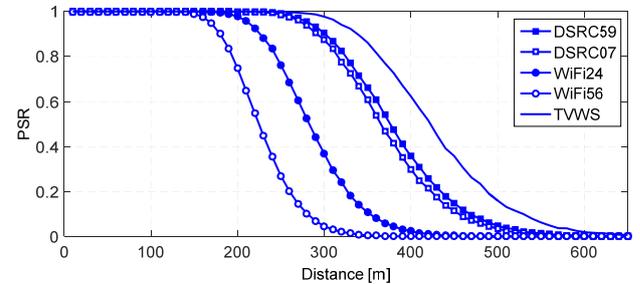
Fig. 2. PSR (Packet Sensing Ratio) for different RATs.

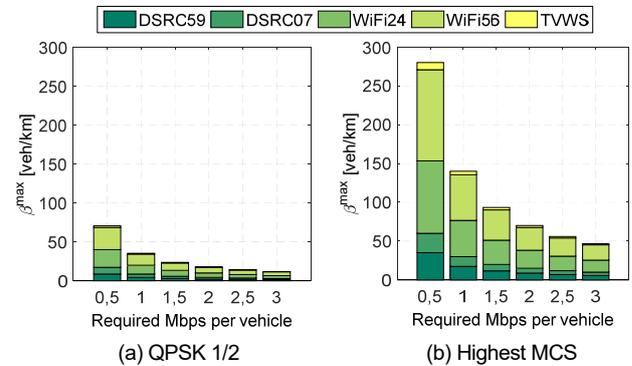

(a) QPSK 1/2      (b) Highest MCS

Fig. 3. Upper-bound of the maximum traffic density supported.

## 4 HETEROGENEOUS V2V PROPOSAL

An heterogeneous V2V communications algorithm should be able to dynamically select for each vehicle the most adequate RAT in order to satisfy its application requirements and maximize the network capacity. Finding the optimum solution to this selection problem can be a chal-lenging task given the large number of possible solutions and the strict latency requirements that generally characterize V2V applications. A scenario with $v$ vehicles and $i$ RATs per vehicle has $i^v$ possible solutions. Even when considering a medium to low density of vehicles (e.g. $v=20$), the number of possible solutions (~$3 \cdot 10^9$) is quite significant with only $i=3$ possible RATs. The proposed heterogeneous V2V communications algorithm (CARHet) reduces this computational cost by taking decisions locally at each vehicle. Each vehicle seeks its local optimal solution taking into account the decisions previously taken by its neighbor vehicles, and the impact that its decision could have on its neighbor vehicles. In particular, each vehicle dynamically selects for its transmissions the RAT that satisfies its application requirements with the minimum cost. The application generates $R$ bps that are transmitted in 1-hop broadcast packets. The application requires that at least $P$% of the transmitted packets are correctly received at distance $D$ (i.e. it requires a throughput higher or equal than $P \cdot R$ at distances lower or equal than $D$). These requirements have been set following the 3GPP guidelines in [3] that specify target minimum reception reliability levels (or PDR, Packet Delivery Ratio) at the established distance. The cost is here measured as the channel load, but other metrics could also be valid. Vehicles implementing CARHet take into account the communications context of neighbor vehicles to select their RAT. To this aim, vehicles periodically share information about the status of their RATs. When a vehicle needs to select a RAT, it estimates the performance it could achieve with every available RAT, the cost (or channel load) it will experience if selecting such RAT, and also the cost that selecting such RAT could generate on neighbor vehicles. Fig. 4 depicts the flow chart of CARHet that could be implemented in the transversal management layer defined in the ITS station reference architecture (Fig. 1). Its main modules are next detailed.

*Context acquisition and Context sharing (Modules I and II)*. With CARHet, vehicles periodically measure and exchange the channel load they sense on all available RATs. More specifically, vehicles estimate the channel load using the CBR and exchange it every $T_{meas}$ using timer $t_m$ in Fig. 4. This information is broadcasted in a CIS (Context Information Sharing) packet using the RAT selected for data transmission. The CIS packet also includes the position of the transmitting vehicle, and the position and channel load measurements of its 1-hop neighbors. The information of the 1-hop neighbors is re-broadcasted so that each vehicle takes into account the context of its 2-hop neighbors when selecting its RAT. This is done because the transmissions of a given vehicle can interfere up to 2 hops [30]. Using received CIS packets, each vehicle creates its own context table that includes the position and channel load measured by its 1-hop and 2-hop neighbors. Table II shows an example of a context table built by a given vehicle A in a scenario with 4 vehicles (A, B, C and D) and 3 RATs (RAT$_1$, RAT$_2$, RAT$_3$). This example assumes that vehicles B and C are 1-hop neighbors of A, and D is a 2-hop neighbor. Vehicle A receives the information of D through vehicle C. To maintain the table updated, every time a vehicle receives a CIS packet, it updates the $RT$ (Reception Time) and the $UT$ (Update Time) parameters in the table. $RT$ represents the time when the last packet was received from a given vehicle, and $UT$ the last time the information was updated for each 1-hop and 2-hop neighbor. $RT$ is equal to $UT$ for 1-hop neighbors, but it is not available for 2-hop neighbors since their context information is received through other vehicles. In Table II, $X$ and $Y$ represent the latitude and longitude of the vehicles. A vehicle is deleted from the table if its information is not updated (directly or indirectly) during the last $T_{neigh}$. The information that vehicle A would transmit in its CIS packets is highlighted in grey color in Table II. It includes its own information and information about its 1-hop neighbors (i.e. vehicles B and C in the example). Vehicle A receives the information about its 1-hop neighbors through their CIS packets. It retransmits this information in its CIS packets so that other vehicles can take it into account when selecting their RATs. Modules I and II describe the context acquisition and sharing processes of CARHet.

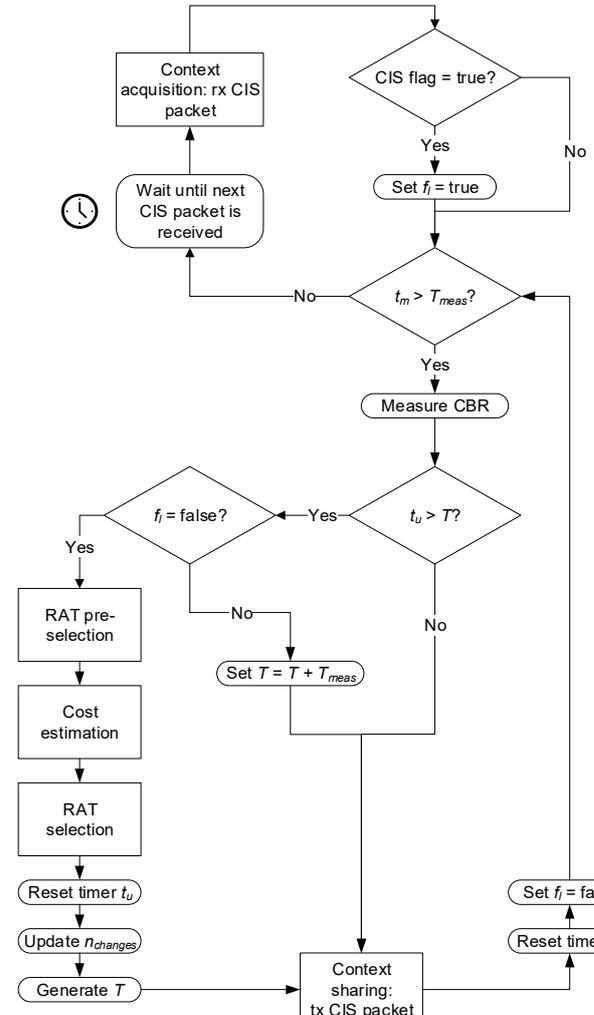

Fig. 4. Flow chart of CARHet.

TABLE II. EXAMPLE OF CONTEXT TABLE

| Vehicle | RT | UT | Position | CBR | | |
| --- | --- | --- | --- | --- | --- | --- |
| | | | | RAT$_1$ | RAT$_2$ | RAT$_3$ |
| A | - | 5.12s | $X_A, Y_A$ | 32% | 5% | 6% |
| B (1-hop) | 5.36s | 5.36s | $X_B, Y_B$ | 20% | 56% | 36% |
| C (1-hop) | 5.27s | 5.27s | $X_C, Y_C$ | 37% | 45% | 35% |
| D (2-hops) | - | 5.27s | $X_D, Y_D$ | 44% | 25% | 24% |

It is important noting that the context acquisition and context sharing modules have been designed so that accurate and updated 2-hop information can be maintained. This is achieved by configuring all vehicles to periodically report: their position, the CBR they measure per RAT, and the position and CBR measurements per RAT of their 1-hop neighbors. This ensures that vehicles always have the necessary information to select the most adequate RAT. In fact, multiple vehicles in the same area will experience and report similar CBR levels. A vehicle implementing CARHet would hence require receiving the information from just one of these vehicles. In addition, multiple vehicles receive (and retransmit) the CBR experienced by a given vehicle. This ensures the necessary redundancy needed by CARHet.

MODULE I. CONTEXT ACQUISITION
Input: CIS packet received from a 1-hop vehicle neighbor
Output: updated context table
Execution: when a CIS packet is received

1. **For** each vehicle $i$ whose data is included in the packet **do**
2.     **If** $UT_i$ received in the CIS>$UT_i$ in the context table **then**
3.        Update $UT_i$ in the table
4.        Update position of vehicle $i$ in the table
5.        **For** each RAT $j$ with $1 \leq j \leq N_{RAT}$ **do**
6.           Update in the table the load in RAT $j$ for vehicle $i$
7.        **End For**
8.     **End if**
9. **End For**

MODULE II. CONTEXT SHARING
Input: context table
Output: CIS packet
Execution: every $T_{meas}$.

1. **For** each 1-hop neighbor $i$ in the table & own vehicle **do**
2.     Add $UT$ of vehicle $i$ to the CIS packet
3.     Add position of vehicle $i$ to the CIS packet
4.     **For** each RAT $j$ with $1 \leq j \leq N_{RAT}$ **do**
5.        Add load in RAT $j$ by vehicle $i$ to the CIS packet
6.     **End For**
7. **End For**

*RAT pre-selection (Module III).* This process is in charge of identifying and pre-selecting the available RATs that can satisfy the application requirements whenever CARHet is executed. In this study, the application requires that at least $P$=90% of the transmitted packets are correctly received at distance $D$. A RAT is hence considered to satisfy the application requirements if the PDR is higher or equal than 0.9 at the distance $D$. This reliability level has been selected following the 3GPP guidelines in [3] where reliability levels (between 85% and 95%) are identified for different scenarios. 90% is considered for the sensor and state map sharing application in [4]. This application enables sharing of raw or processed sensor data to build collective situational awareness. The PDR is influenced by the channel load and interference. We have hence derived PDR curves for each RAT for different CBR levels[2]. Fig. 5 represents a PDR example for

[2] The PDR curves are obtained using the simulator presented in Section 5.

DSRC at 5.9GHz and CBR levels varying between 0 and 0.9. Similar curves have been derived for all the implemented RATs. The RAT pre-selection process works as follows. If a vehicle has to select a RAT, it will measure the CBR experienced in all available RATs. For each RAT and experienced CBR level, the vehicle derives the PDR at distance $D$. CARHet then pre-selects those RATs that are capable to satisfy a PDR equal or higher than 0.9 at distance $D$.

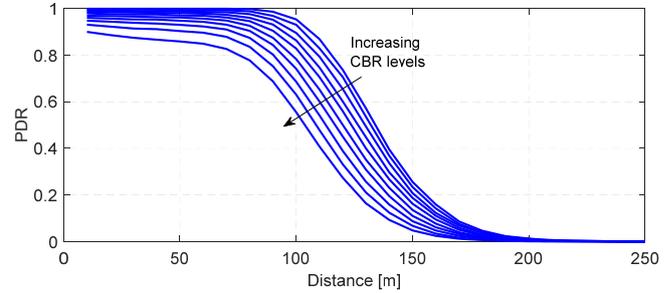

Fig. 5. PDR (Packet Delivery Ratio) for DSRC at 5.9GHz for average CBR levels varying between 0 and 0.9 in steps of 0.1.

MODULE III. RAT PRE-SELECTION
Inputs: $D$, $R$ and PDR curves for the current CBR
Output: each RAT is pre-selected or not as candidate RAT
Execution: every $T_{update}$

1. **For** each RAT $j$ with $1 \leq j \leq N_{RAT}$ **do**
2.     **If** $PDR_j(D)$>0.9 **then**
3.        Pre-select RAT $j$ as candidate RAT
4.     **End If**
5. **End For**

*Cost estimation (Module IV).* CARHet computes then the cost associated to the use of each pre-selected RAT that is able to satisfy the application requirements. In this study, the cost is measured as the CBR that a RAT would experience if it is selected by the vehicle that is executing CARHet. The cost of using a given RAT $j$ is equal to the maximum CBR that would be experienced by its 1-hop and 2-hop neighbors. This cost is represented by $c_j$=max$\{L_{ij}\}$ with $L_{ij}$ representing the CBR that would be experienced by neighbor $i$ if RAT $j$ is selected. To compute the cost, the vehicle needs to estimate $L_{ij}$ for each one of its 1-hop and 2-hop neighbors and all RATs as specified in Module IV. $L_{ij}$ can be computed as:

$$L_{ij} = LE_{ij} + LG_{ij} \qquad (5)$$

where $LE_{ij}$ is the CBR experienced by neighbor $i$ with RAT $j$, and $LG_{ij}$ is the additional CBR the vehicle executing CARHet would generate to neighbor $i$ if RAT $j$ is selected. $LE_{ij}$ is measured by neighbor $i$ and is included in its CIS packets; the information is hence stored in the context table (Table II). $LG_{ij}$ can be estimated as follows:

$$LG_{ij} = n \cdot t_j \cdot PSR_j(d_i) \qquad (6)$$

where $n$ represents the number of packets generated per second, $t_j$ the packet duration, $d_i$ the distance between the transmitting vehicle and vehicle $i$, and $PSR_j(d_i)$ the packet sensing ratio at distance $d_i$ for RAT $j$.

MODULE IV. COST ESTIMATION
Inputs: $D$, $R$, context table and PSR
Output: $c_j$
Execution: every $T_{update}$.

1. **For** each RAT $j$ with $1 \leq j \leq N_{RAT}$ **do**
2.    **If** RAT $j$ was pre-selected in Module III **then**
3.       Initialize the maximum channel load $c_j$ as 0
4.       **For** each 1-hop and 2-hop vehicle neighbors $i$ **do**
5.          Compute $LG_{ij}$ using equation (7)
6.          Extract $LE_{ij}$ from the context table
7.          Compute $L_{ij}$ using equation (6)
8.          **If** $L_{ij} > c_j$ **then**
9.             Set $c_j$ equal to $L_{ij}$
10.          **End If**
11.       **End For**
12.    **Else**
13.       Set the maximum channel load $c_j$ as 100%
14.    **End if**
15. **End For**

*RAT selection (Module V)*. The RAT selection process identifies the pre-selected RAT that minimizes the maximum channel load $c_j$. The process computes then the difference between the maximum load experienced with the pre-selected RATs and with the one currently utilized by the vehicle executing CARHet. The RAT is only changed if this difference is higher than a threshold, $\alpha$, to avoid RAT oscillations when the load improvement is minimal.

MODULE V. RAT SELECTION
Inputs: $c_j$ for each RAT
Output: selected RAT
Execution: every $T_{update}$.

1. Initialize $c$ as 100%
2. **For** each RAT $j$ with $1 \leq j \leq N_{RAT}$ **do**
3.    **If** $c_j < c - \alpha$ **then**
4.       Set $c$ equal to $c_j$
5.       Set RAT $j$ as the selected RAT for data transmission
6.    **End if**
7. **End For**

*Decision sharing*. Multiple vehicles can take the same decision (i.e. select the same RAT) if they execute CARHet around the same time. This circumstance could generate instability if all vehicles try to reduce the load of a certain RAT simultaneously. In this case, they could overload a different RAT, and require quickly changing the RAT again. To address this problem, CARHet requires vehicles changing their RAT to inform nearby vehicle by including the CIS flag in the next CIS packet they broadcast. CARHet propagates the CIS flag up to two hops (always attached to CIS packets) since we assume that the load generated by a vehicle affects vehicles up to two hops. All vehicles receiving this information (active CIS flag) postpone the RAT selection process by $T_{meas}$. To do so, the variable $f_l$ is used in Fig. 4.

*CARHet triggering*. The RAT selection process is executed every $T$ seconds in this study (proactive approach) using timer $t_u$ (see Fig. 4). $T$ is a random variable uniformly distributed between $T_{update}$ and $T_{update} \cdot (n_{changes}+1)$. $T_{update}$ is a constant parameter that is common to all vehicles. $n_{changes}$ is the number of consecutive RAT changes performed by a vehicle. This randomization reduces the probability to produce an instable situation where multiple vehicles re-evaluate (and maybe change) their RAT nearly at the same time. This situation could still be produced if a CIS packet containing an active CIS flag is lost due to propagation or interference. To combat instabilities, the length of the randomization interval increases if the instability augments. This is the case because the interval is a function of $n_{changes}$.

## 5 SIMULATION SCENARIOS AND SETTINGS

CARHet has been evaluated using VEINS (Vehicles in Network Simulation), an open source framework for vehicular network simulations that utilizes OMNeT++ and SUMO (Simulation of Urban MObility). A highway traffic scenario with 4 lanes (2 lanes per driving direction) has been simulated using mobility patterns generated by SUMO. Vehicles move at a maximum speed of 100km/h. Different traffic densities are simulated: 40, 80 and 120 veh/km. Each vehicle is equipped with 5 RATs (Table I) that can be simultaneously used: DSRC (IEEE 802.11p) operating at 5.9GHz, DSRC (IEEE 802.11p) operating at 700MHz, WiFi operating at 5.6GHz, WiFi operating at 2.4GHz, and an OFDM-like technology operating in the TVWS band at 460MHz. A vehicle can transmit data using one RAT and simultaneously receive information through all available RATs. These technologies have only been selected for the purpose of illustrating the potential of heterogeneous V2V communications in multi-link and multi-RAT scenarios. The propagation conditions are modeled using the Winner+ B1 model previously described and that has been implemented in VEINS.

Two application scenarios have been simulated in this study. In both scenarios, vehicles periodically broadcast packets of 1024 bytes and the applications require to correctly receive 90% of the transmitted packets at $D$. In the first scenario, $D$ is set equal to 40m, and all vehicles in the scenario transmit the same amount of information $R$ bps. Simulations have been done for $R$ equal to 0.5Mbps, 1Mbps and 1.5Mbps. These data rates are representative of connected and automated vehicular applications. For example, [4] establishes a 0.5Mbps data rate for the "Information sharing for partial/conditional automated driving" use case. This use case requires vehicles sharing detected objects with neighboring vehicles. In the second scenario, 50% of the vehicles are configured with $R$=1.5Mbps and $D$=40m, 25% of the vehicles with $R$=1.0Mbps and $D$=80m, and the remaining 25% of vehicles with $R$=0.5Mbps and $D$=120m. This scenario has been chosen to emulate cooperative perception or sensing applications where vehicles need to exchange more (or richer) sensor data (and therefore need higher throughput) with vehicles at shorter distances than with vehicles at large distances. Shorter distances entail higher risks, and hence a more accurate perception of the environment is required. Higher data rates allow exchanging more sensor data and hence build a more accurate view of the environment.

CARHet is compared in this study to a technique that randomly selects the RAT of each vehicle every $T_{update}$ (to the authors' knowledge, no other reference schemes are available in the literature). Randomly selecting the RAT distributes the vehicles among the different technologies, has very low computational complexity, and does not require any signaling. The results are also compared to the case in which vehicles only utilize IEEE 802.11p at 5.9GHz in order to highlight the current limitations to support connected automated vehicles and the need for heterogeneous V2V communications. Table IV presents the main simulation parameters, including the configuration values of CARHet. Relatively low values for $T_{meas}$ and $T_{update}$ have been selected so that CARHet can quickly react to changing communication context conditions. Larger values would reduce the frequency of RAT changes, but would also result in vehicles not using the best RAT for longer periods of time.

TABLE IV. SIMULATION CONDITIONS

| | Parameter | Values |
|---|---|---|
| Scenario | Highway length [km] | 3 |
| | Traffic density [veh/km] | 40, 80, 120 |
| | Number of lanes | 4 (2 in each direction) |
| | Maximum speed [km/h] | 100 |
| | Simulation time [s] | 250 |
| CARHet | $T_{meas}$ [s] | 0.2 |
| | $T_{update}$ [s] and $T_{neigh}$ [s] | 1 |
| | $\alpha$ | 5% |

## 6 EVALUATION

Fig. 6 depicts the CBR and throughput experienced per vehicle when all vehicles use DSRC at 5.9GHz and have the same application requirements (transmit $R$=0.5Mbps and require that at least $P$=90% of the transmitted packets are correctly received within $D$=40m). The results are presented using box plots, which are widely used in descriptive statistics to graphically depict groups of numerical data. In each box plot, the top and bottom of the box are the 25th and 75th percentiles and therefore the distance between them is the interquartile range. The red horizontal line inside the box represents the median. The whiskers are lines extending above and below each box and represent the 5th and 95th percentiles. Fig. 6a highlights the saturation of IEEE 802.11p as the CBR exceeds the recommended value of 0.6 [28] for all traffic densities. These results are in line with the estimations in Fig. 3b that indicated that the maximum traffic density supported by IEEE 802.11p if all vehicles transmit 0.5Mbps is 35 veh/km. Fig. 6b shows that only for a traffic density of 40 veh/km, vehicles can satisfy throughput values around $R$=0.5Mbps at distances lower than $D$=40m. IEEE 802.11p cannot satisfy the application requirements if the traffic density or the value of $R$ increase.

Fig. 7 and Fig. 8 plot the CBR and throughput, respectively, when all vehicles require $D$=40m and $R$=1.0Mbps, and randomly select the RAT or implement CARHet. Fig. 7a shows that a random (and hence uniform) distribution of vehicles between RATs results in a different channel load per RAT since each RAT has different communication parameters (Table I), in particular different bandwidth. This unequal distribution of the channel load among RATs results in a very different throughput per vehicle, with the differences increasing with the traffic density (Fig. 8a). This results in that there is a significant percentage of vehicles that cannot satisfy the application requirements when they randomly select their RAT (Fig. 9). A vehicle is considered to be satisfied if its throughput is equal or higher than $0.9 \cdot R$ at distances equal and lower than $D$=40m.

Fig. 7b shows that CARHet is capable to balance the channel load among RATs despite their different characteristics. This is particularly noticeable when comparing the median of the CBR[3]. A more balanced channel usage among RATs results in significantly higher (and more homogeneous) throughput values per vehicle with CARHet (Fig. 8b) compared to the case in which vehicles randomly select their RAT (Fig. 8a). This results in a significantly higher percentage of vehicles satisfied when implementing CARHet compared to when randomly selecting the RAT (Fig. 9). A com-

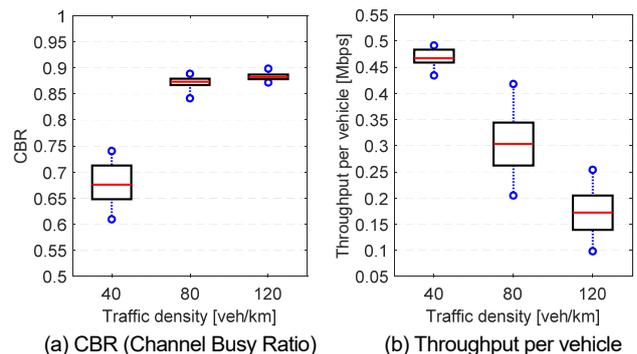
(a) CBR (Channel Busy Ratio)  (b) Throughput per vehicle
Fig. 6. CBR and throughput per vehicle when all vehicles use DSRC at 5.9GHz and they require $R$=0.5Mbps and $D$=40m.

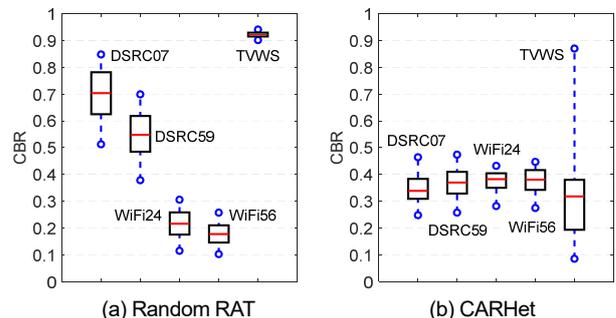
(a) Random RAT  (b) CARHet
Fig. 7. CBR when vehicles randomly select a RAT or implement CARHet. All vehicles require $R$=1.0Mbps and $D$=40m. Traffic density: 80 veh/km.

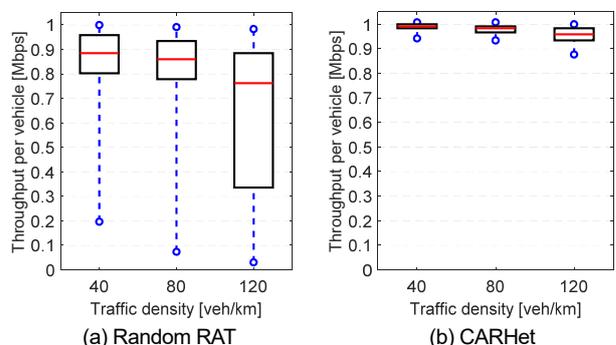
(a) Random RAT  (b) CARHet
Fig. 8. Throughput per vehicle when vehicles randomly select a RAT or implement CARHet when all vehicles require $R$=1.0Mbps and $D$=40m.

[3] Vehicles moving in opposite directions result in changes of the channel load over space and time. The variations increase as the bandwidth and data rate decrease (TVWS is the most affected RAT), which explains the box plot differences in Fig. 7b.

parison of Fig. 9 and Fig. 3b shows that CARHet can approximate the maximum traffic densities estimated in Section 3. For example, Fig. 3b estimated the maximum traffic density for $R$=1.0Mbps to be equal to approximately 140veh/km. Fig. 9b shows that CARHet approaches this maximum capacity as it can satisfy approximately 90% of the vehicles when the traffic density is equal to 120veh/km. The maximum traffic density estimated for $R$=1.5Mbps was approximately 90veh/km (Fig. 3b). Fig. 9c shows that CARHet can satisfy more than 90% of vehicles for 80veh/km, but the percentage of vehicles satisfied decreases for 120veh/km.

The RAT selection algorithms (Random and CARHet) are executed by each vehicle every $T_{update}$=1s. This value was chosen so that the selection process can adequately follow relevant changes in the communication context conditions. Fig. 10 shows that CARHet guarantees a stable operation that prevents vehicles constantly changing the RAT if such change has little impact on the capacity to satisfy the application requirements. Fig. 10 depicts the time between RAT changes per vehicle (τ) when vehicles randomly select the RAT every $T_{update}$ (Fig. 10a) and when they implement CARHet (Fig. 10b). With the random scheme, the probability that a vehicle changes its RAT is equal to 4/5. This is equivalent to approximately changing the RAT every 1.25s. CARHet significantly reduces the number of RAT changes per second per vehicle[4] as vehicles tend to change the RAT every 50-80s on average (Fig. 10b). The results in Fig. 10b and Fig. 9 suggest that CARHet is capable to limit the RAT changes to those that have a positive impact on the capacity to satisfy the application requirements.

The stability and convergence of CARHet is also illustrated with Fig. 11. The figure plots the time evolution of the CBR measured by a vehicle during 10 seconds. The vehicle has been randomly selected in the scenario. The Random algorithm (Fig. 11a) converges to a solution with unequal distribution of the channel load among RATs (also observed in Fig. 7a). On the other hand, CARHet (Fig. 11b) converges to a solution that is capable to balance the load among RATs (also observed in Fig. 7b). Fig. 11 also shows that CARHet converges to a stable solution. This is actually achieved even if during 10 seconds more than half of the neighbors of a vehicle change in the considered scenario[5]. We would like to highlight that the trends observed in Fig. 11 are maintained for different time windows and randomly selected vehicles. Fig. 11 is an example to illustrate the capacity of CARHet to converge to a stable solution. In fact, the solution reached by CARHet results in that each vehicle transmits using the RAT that minimizes the maximum channel load experienced by vehicles up to 2 hops. It should be noted that it is not possible to guarantee the same channel load for all RATs since the channel load depends on the propagation conditions, the characteristics and configuration of each RAT, and the mobility of vehicles. In fact, vehicles at different locations experience different channel load levels from the same transmitting vehicle. In addition, each RAT has a different bandwidth and transmission parameters. It is then difficult that a given number of vehicles using a given RAT generate exactly the same channel load that a different number of vehicles using another RAT. Finally, it should also be noted that the mobility of vehicles results in channel load variations even if all vehicles maintain their RAT selection and configuration.

The previous results were obtained when all the vehicles require the same $R$ and $D$. Fig. 12 depicts the throughput per vehicle obtained when vehicles have different application requirements as detailed earlier in this section. In this case, 50% of the vehicles are configured with $R$=1.5Mbps and $D$=40m, 25% of the vehicles with $R$=1.0Mbps and $D$=80m, and the remaining 25% of the vehicles are configured with $R$=0.5Mbps and $D$=120m. Fig. 12a shows that a random selection of the RAT results in that a non-negligible percentage of vehicles experience a throughput significantly lower than the throughput demanded by the application. In this scenario, this result is not only due to the fact that randomly selecting the RAT can overload certain channels, but also to the fact that not all RATs can satisfy the application requirements. The throughput performance depicted in Fig. 12a is at the origin of the low percentage of vehicles satisfied when randomly selecting the RAT, shown in Fig. 13. Fig. 13 shows that CARHet is capable to satisfy a significant percentage of vehicles also in the scenarios where vehicles have mixed application requirements. The higher satisfaction levels obtained with CARHet result from the fact that CARHet is capable to match vehicles with the RATs that are capable to satisfy their application requirements. This results in the higher average throughput values per vehicle experienced with CARHet and its lower throughput interquartile range (Fig. 12b). This low range indicates that CARHet provides similar QoS levels to the majority of vehicles.

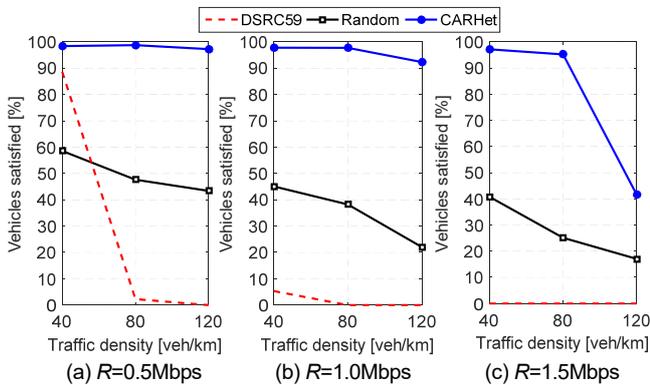

Fig. 9. Percentage of vehicles satisfied when all vehicles require $D$=40m.

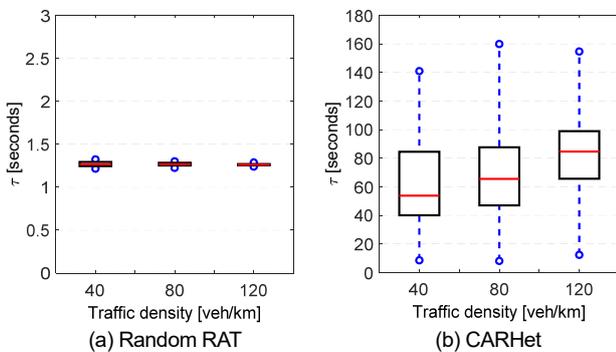

Fig. 10. Time between RAT changes per vehicle (τ) in seconds when all vehicles require $R$=1.0Mbps and $D$=40m.

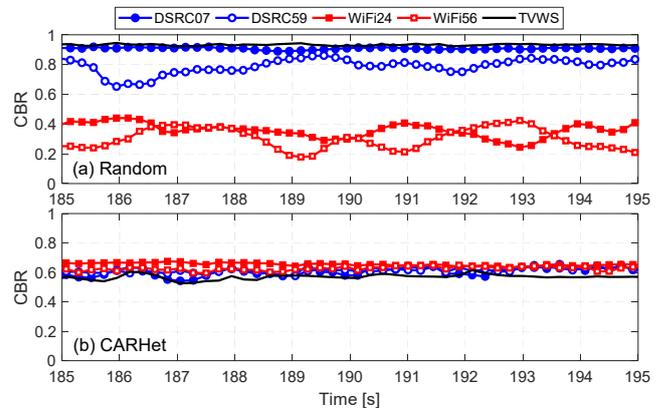

Fig. 11. Time evolution of the CBR (Channel Busy Ratio) measured by a randomly selected vehicle. All vehicles in the scenario require $R$=1.0Mbps and $D$=40m. The traffic density is 120 veh/km.

---

[4] RAT changes are executed at the vehicle level with no additional signaling required at the network level.

[5] Vehicles move at a maximum speed of 100km/h or 27.7m/s and travel in the two driving directions.

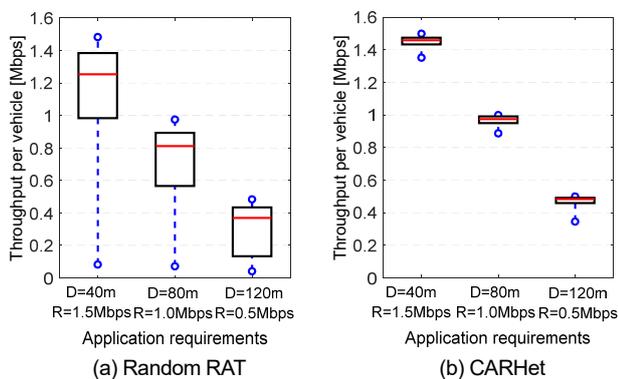

Fig. 12. Throughput per vehicle when vehicles randomly select a RAT or implement CARHet. The throughput is shown as a function of the application requirements. The scenario considers mixed application requirements and a traffic density of 80veh/km.

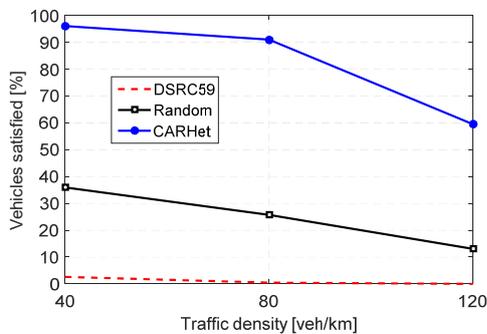

Fig. 13. Percentage of vehicles satisfied when vehicles have mixed application requirements.

## 7 COMPUTATIONAL COST AND COMMUNICATION OVERHEAD

This section analyzes the computational cost and communication overhead of CARHet, and hence its feasibility. Table V reports the number of CPU cycles needed to execute CARHet. The information is presented separately for each one of the CARHet modules detailed in Section 4. Each term in the sums corresponds to the number of cycles needed to execute each line of the modules' pseudo-code. The values shown in Table V correspond to upper bounds since they have been estimated considering that all the conditions evaluated in Module I to Module V are met, and hence the instructions inside the *for* or *if* loops are executed. The number of CPU cycles needed to execute CARHet depends on the number of RATs available at each vehicle ($N_{RAT}$), the number of vehicle neighbors at 1 hop ($N_1$), and the number of vehicle neighbors at 2 hops ($N_2$). It also depends on the $T_{meas}$ and $T_{update}$ parameters since these parameters influence how often CARHet is executed and how often CIS packets are transmitted. The number of CPU cycles has been computed considering Intel CPU architectures [31]. In this case, the multiplication of two floating point numbers requires 5 CPU cycles, and their addition requires 3 cycles. Fig. 14 shows an example of the impact that executing CARHet will have on the CPU of a vehicle. The figure plots an upper-bound of the amount of CPU usage (or percentage of the CPU's capacity) consumed by CARHet for different CPU speeds and number of neighbors. The figure has been derived considering $N_{RAT}$=5 and $N_1$=$N_2$=$N$ (i.e. each vehicle has the same number of 1-hop and 2-hop neighbors). Fig. 14 shows that the CPU usage is less than 0.3%, which demonstrates the low computational requirements of CARHet.

TABLE V. CARHet COMPUTATIONAL COST PER VEHICLE

| Module | Execution freq. (Hz) | Number of CPU cycles |
|---|---|---|
| I. Context acquisition | $N_1/T_{meas}$ | $2 \cdot N_1 + N_1 + 2 \cdot N_1 + 2N_1 \cdot N_{RAT} + N_1 \cdot N_{RAT}$ |
| II. Context sharing | $1/T_{meas}$ | $2 \cdot N_1 + N_1 + 2 \cdot N_1 + 2 \cdot N_1 \cdot N_{RAT} + N_1 \cdot N_{RAT}$ |
| III. RAT pre-selection | $1/T_{update}$ | $2 \cdot N_{RAT} + 2 \cdot N_{RAT} + N_{RAT}$ |
| IV. Cost estimation | $1/T_{update}$ | $2 \cdot N_{RAT} + N_{RAT} + N_{RAT} + 2 \cdot N_{RAT} \cdot (N_1 + N_2) +$ $11 \cdot N_{RAT} \cdot (N_1 + N_2) + N_{RAT} \cdot (N_1 + N_2) + 3 \cdot N_{RAT}$ $\cdot (N_1 + N_2) + N_{RAT} \cdot (N_1 + N_2) + N_{RAT} \cdot (N_1 + N_2)$ |
| V. RAT selection | $1/T_{update}$ | $1 + 2 \cdot N_{RAT} + N_{RAT} + N_{RAT} + N_{RAT}$ |

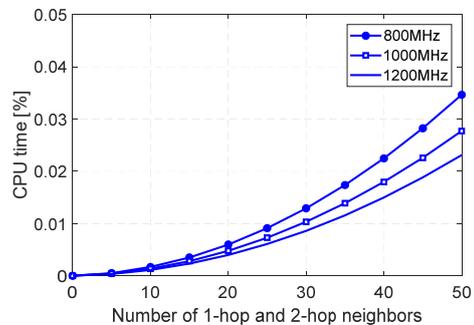

Fig. 14. Upper-bound of CARHet's CPU usage for different processor speeds, and considering $N_{RAT}$=5 and $N_1$=$N_2$=$N$.

CARHet requires vehicles to exchange context information using the CIS packets. These packets represent then CARHet's communication cost or overhead. To quantify this overhead, we need to take into account that each vehicle transmits a CIS packet every $T_{meas}$. A vehicle includes in a CIS packet: 1) its position, 2) the channel load it has measured for each RAT and the timestamp of these measurements, and 3) the position of its 1-hop neighbors and their channel load measurements per RAT. The overhead $O_v$ generated by CARHet can be estimated in bits per second as:

$$O_v = N_{CIS} \cdot s_{CIS} \quad (7)$$

where $N_{CIS}$ is the number of CIS packets received per second per vehicle, and $s_{CIS}$ represents the size in bits of a CIS packet. If we consider $N$ 1-hop and 2-hop neighbors, $N_{CIS}$ can be upper-bounded by $N_{CIS}=N/T_{meas}$ and $s_{CIS}$ by:

$$s_{CIS} = (s_T + s_{Lat} + s_{Lon} + N_{RAT} \cdot s_{CBR}) \cdot (N+1) \qquad (8)$$

where $s_T$, $s_{Lat}$, $s_{Lon}$ and $s_{CBR}$ represent the size in bits of the timestamp, latitude, longitude and CBR fields, respectively. We consider that the timestamp, latitude and longitude fields are encoded with 4 bytes each[6], and each CBR value with 1 byte (there is one CBR value per RAT). Fig. 15 plots CARHet's overhead ($O_v$) as a function of the number of 1-hop and 2-hop neigbhors, $N$. The overhead is normalized by the sum of the bandwidth of all RATs (i.e. 66MHz, see Table I) so that it is expressed in b/s/Hz. Fig. 15 shows that the upper-bound of the overhead generated by CARHet (or its communication cost) is quite low, even for the highest number of neighbors[7]. This overhead is further reduced when $T_{meas}$ increases. However, increasing $T_{meas}$ reduces the update rate of the context information, which can have an impact on CARHet's capacity to rapidly adapt to changing communication conditions.

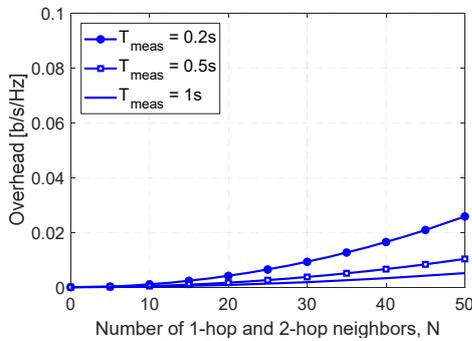

Fig. 15. Upper-bound of CARHet's overhead (communication cost) considering $N_{RAT}$=5 and $N_1$=$N_2$=$N$.

## 8 CONCLUSIONS AND DISCUSSION

A widespread deployment of connected vehicles and the introduction of connected automated driving applications will notably increase the bandwidth and scalability requirements of vehicular networks. This paper proposes to address this challenge by adopting heterogeneous networking for V2X communications in multi-link and multi-RAT vehicular scenarios. In particular, this paper proposes and evaluates CARHet, a novel context-aware heterogeneous V2V communications algorithm that allows each vehicle to autonomously and dynamically select its communication technology (or RAT) based on its application requirements and the communication context conditions observed by neighboring vehicles. To the author's knowledge, this is the first heterogeneous V2V communications algorithm proposed in the literature that is technology and application agnostic, and that allows each vehicle to autonomously and dynamically select the communications technology for its V2V transmissions. CARHet has been evaluated considering a given set of communications technologies for illustration purposes. However, it could well be extended to consider other RATs.

The conducted study has demonstrated that heterogeneous V2V communications can help address the bandwidth and scalability requirements that future vehicular networks will face. The study has also shown that CARHet is capable to adequately distribute the load among RATs, and ensure high and homogenous QoS levels across the network with a low computational and communication cost. As a result, CARHet can satisfy the application requirements for a large percentage of vehicles while approximating the estimated upper bound of the network capacity.

CARHet is a first proposal towards the design of future heterogeneous V2V solutions, with still many contributions to be expected from the community. For example, heterogeneous V2V algorithms can be designed with other objectives in mind (e.g. reliability rather than scalability), and hence with different performance and cost functions. Solutions will need to be proposed for scenarios in which all vehicles do not have the same RATs[8]. This paper has evaluated a proactive implementation of CARHet where the RAT selection process is triggered periodically. However, reactive or hybrid implementations would also be possible, which could reduce the number of RAT changes by limiting them to situations in which the communication conditions change.

## ACKNOWLEDGMENT

The authors acknowledge the support of the Spanish Ministry of Economy and Competitiveness and FEDER funds (projects TEC2014-57146-R and TEC2017-88612-R).

---

[6] In line with corresponding data elements in ETSI V2X messages.
[7] 50 neighbors is the average number of neighbors observed for the highest traffic density simulated in this study (120 veh/km).

[8] Similar challenges can be foreseen once first V2X technologies are deployed (e.g. DSRC based on IEEE 802.11p) and new V2X standards (e.g. 5G V2X) are proposed for increasing the communications capabilities and enable additional functionality. In this case, new vehicles will have more V2X technologies on board than older vehicles.

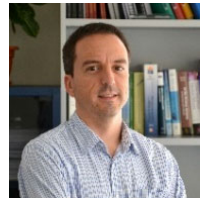

**Miguel Sepulcre** (msepulcre@umh.es) received a Telecommunications Engineering degree in 2004 and a Ph.D. in Communications Technologies in 2010, both from Universidad Miguel Hernández de Elche (UMH), Spain. He was awarded by the COIT (Spanish association of Telecomms. Engineers) with the prize to the best Ph.D. thesis. He has been visiting researcher at ESA (The Netherlands), at Karlsruhe Institute of Technology (Germany), and at Toyota InfoTechnology Center (Japan). He serves as Associate Editor for IEEE Vehicular Technology Magazine and IEEE Communications Letters. He was TPC Co-Chair of IEEE VTC2018-Fall, IEEE/IFIP WONS 2018 and IEEE VNC 2016. He is Assistant Professor at UMH, and member of UWICORE research lab working in wireless vehicular networks and industrial wireless networks.

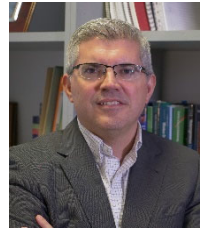

**Javier Gozalvez** (j.gozalvez@umh.es) received an electronics engineering degree from the Engineering School ENSEIRB (Bordeaux, France), and a PhD in mobile communications from the University of Strathclyde, Glasgow, U.K. Since October 2002, he is with the Universidad Miguel Hernández de Elche (UMH), Spain, where he is currently a Full Professor and Direc-tor of the UWICORE laboratory. At UWICORE, he leads research activities in the areas of vehicular networks, 5G and Beyond device-centric wireless networks, and industrial wireless networks. He is an elected member to the Board of Governors of the IEEE Vehicular Technology Society (VTS) since 2011, and served as President of the IEEE IEEE VTS in 2016 and 2017. He was an IEEE Distinguished Lecturer for the IEEE VTS, and currently serves as IEEE Distinguished Speaker. He is the Editor in Chief of the IEEE Vehicular Technology Magazine. He is the General Co-Chair for the IEEE Connected and Automated Vehicles Symposium 2019 and 2018, and was the General Co-Chair for the IEEE VTC-Spring 2015 conference in Glasgow (UK).